\documentclass[conference, 10 pt]{IEEEtran}

\usepackage{soul} 
\IEEEoverridecommandlockouts
\usepackage{cite}
\usepackage{amsmath,amssymb,amsfonts}
\usepackage{braket}
\usepackage{algorithmic}
\usepackage{multirow}
\usepackage{graphicx}
\usepackage{textcomp}
\usepackage{xcolor}
\def\BibTeX{{\rm B\kern-.05em{\sc i\kern-.025em b}\kern-.08em
    T\kern-.1667em\lower.7ex\hbox{E}\kern-.125emX}}
\begin{document}

\title{Optimization of Quantum Error Correcting Code under Temporal Variation of Qubit Quality\\
}

\author{\IEEEauthorblockN{Subrata Das}
\IEEEauthorblockA{\textit{Dept. of Electrical Engineering} \\
\textit{The Pennsylvania State University}\\
University Park, PA, USA \\
sjd6366@psu.edu}
\and
\IEEEauthorblockN{Swaroop Ghosh}
\IEEEauthorblockA{\textit{Dept. of Electrical Engineering} \\
\textit{The Pennsylvania State University}\\
University Park, PA, USA \\
szg212@psu.edu}

}

\maketitle

\begin{abstract}
Error rates in current noisy quantum hardware are not static; they vary over time and across qubits. This temporal and spatial variation challenges the effectiveness of fixed-distance quantum error correction (QEC) codes. In this paper, we analyze 12 days of calibration data from IBM's 127-qubit device (\texttt{ibm\_kyiv}), showing the fluctuation of Pauli-X and CNOT gate error rates. We demonstrate that fixed-distance QEC can either underperform or lead to excessive overhead, depending on the selected qubit and the error rate of the day. We then propose a simple adaptive QEC approach that selects an appropriate code distance per qubit, based on daily error rates. Using logical error rate modeling, we identify qubits that cannot be used and qubits that can be recovered with minimal resources. Our method avoids unnecessary resource overhead by excluding outlier qubits and tailoring code distances. 
Across 12 calibration days on \texttt{ibm\_kyiv}, our adaptive strategy reduces physical qubit overhead by over 50\% per logical qubit while maintaining access to 85--100\% of usable qubits. To further validate the method, we repeat the experiment on two additional 127-qubit devices, \texttt{ibm\_brisbane} and \texttt{ibm\_sherbrooke}, where the overhead savings reach up to 71\% while still preserving over 80\% qubit usability.
This approach offers a practical and efficient path forward for Noisy Intermediate-Scale Quantum (NISQ)-era QEC strategies. 

\end{abstract}


\section{Introduction}
\label{sec:intro}

Quantum error correction (QEC) is essential for achieving reliable quantum computation. In Noisy Intermediate-Scale Quantum (NISQ) devices, physical qubits are subject to various sources of error, including decoherence, control imperfections, and measurement inaccuracy. These error processes, often modeled using Pauli channels or amplitude damping noise, limit the depth and accuracy of quantum circuits. Without error correction, scaling up quantum algorithms remains infeasible. To mitigate such limitations, QEC encodes a logical qubit using multiple physical qubits and applies a decoding strategy that can detect and correct errors arising during computation \cite{gottesman2010, preskill2018}.

\begin{figure}[ht]
    \centering
    \includegraphics[width=0.8\linewidth]{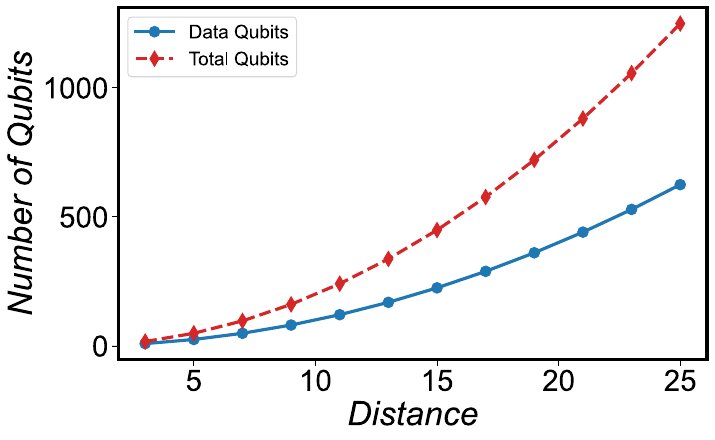}
    \caption{Qubit requirement with increasing code distance. 
    As the distance of the code increases to improve error tolerance, the number of qubits required for both data and ancilla qubits rises, with ancilla qubits contributing significantly to the total overhead \cite{chatterjee2024quantum}.}
    \label{fig:overhead}
\end{figure}

Among the most studied QEC codes, surface codes are favored for their high error threshold (around 1\%) and compatibility with nearest-neighbor qubit connectivity in superconducting platforms \cite{fowler2012, wang2011}. These codes use repeated stabilizer measurements to protect quantum information. The strength of a surface code is determined by its code distance \( d \), which dictates the number of physical errors the code can tolerate. 
Increasing \( d \) reduces the logical error rate exponentially but comes with a steep overhead in terms of physical qubits and gate operations \cite{dennis2002, terhal2015}. This overhead arises from the structure of surface codes, which require not only data qubits to store the logical quantum information but also ancilla qubits to perform stabilizer measurements. Ancilla qubits are used to extract error syndromes without collapsing the encoded quantum state. For each stabilizer operator, one or more ancilla qubits are assigned to interact with neighboring data qubits and report on potential errors. As the code distance increases, both the number of data qubits and the required ancilla qubits grow quadratically. This tradeoff is illustrated in Fig. \ref{fig:overhead}, which shows the total number of qubits required as a function of code distance \cite{chatterjee2024quantum}. The total qubit count rises significantly faster than the number of data qubits alone, due to the growing number of ancilla. For a given distance, the number of ancilla qubits is approximately equal to the number of data qubits, effectively doubling the physical qubit requirement and presenting a major constraint for hardware scalability.

In most current methodologies, a fixed code distance is selected based on average or expected physical error rates and used uniformly across all qubits and times \cite{cross2019}. This practice assumes that error rates are both spatially and temporally homogeneous. However, recent studies and calibration data from real quantum processors reveal significant deviations from this assumption. Qubit-specific error rates vary due to differences in fabrication quality, local crosstalk, and thermal stability \cite{krinner2019}. Moreover, error rates for the same qubit can drift over time due to hardware aging, fluctuation in control electronics, or recalibration cycles \cite{gambetta2020, mckay2019}. For example, IBM’s superconducting devices exhibit measurable day-to-day variation in both single-qubit and two-qubit gate errors \cite{tannu2019not}.

Applying a uniform QEC configuration under such variable conditions introduces inefficiencies. A very low code distance applied to a temporarily noisy qubit can lead to higher logical errors making the error correction useless whereas a very high code distance applied to a resilient qubit can consume excessive physical resources unnecessarily. 
Both outcomes are limiting factors for running useful quantum circuits on near-term devices \cite{corcoles2020}.


This paper presents a strategy for QEC on NISQ hardware while accounting for real-time and per-qubit error rate variations. By leveraging historical calibration data, we demonstrate that selecting code distance adaptively based on current qubit quality enables more efficient use of hardware. Our methodology allows the system to exclude qubits that are not recoverable within a reasonable resource budget and to reduce the overhead for stable qubits that do not require high-distance encoding. The resulting scheme optimizes total qubit overhead while preserving the target logical error rates. We validate our approach using calibration data from three real IBM quantum processors—\texttt{ibm\_kyiv}, \texttt{ibm\_brisbane}, and \texttt{ibm\_sherbrooke}, showing consistent improvements across devices. This work supports the use of dynamic, resource-aware QEC frameworks that respond to hardware variability in actual quantum systems.


In the the rest of the paper, Section II provides background on surface codes and prior work on adaptive quantum error correction. Section III presents the variation in single-qubit and two-qubit error rates observed in the \texttt{ibm\_kyiv} device over 12 days. 
Section IV explains our proposed QEC optimization strategy. 
Section V concludes the paper.

\section{Surface Codes and Adaptive QEC}

\subsection{Fundamentals of Surface Codes}

\begin{figure}[ht]
    \centering
    \includegraphics[width=0.8\linewidth]{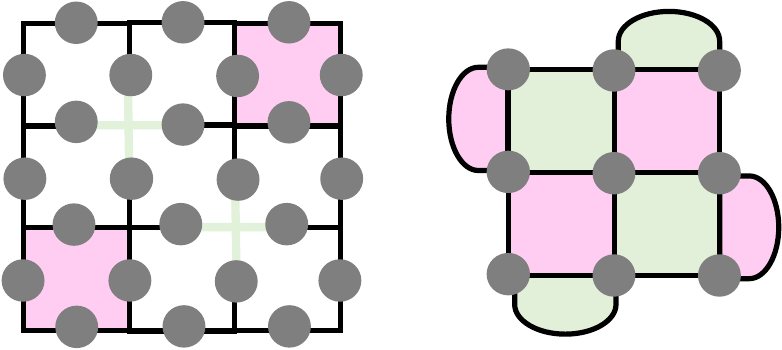}
    \caption{Representation of distance-3 surface codes. Unrotated surface code (left) and rotated
surface code (right). The gray blobs depicting qubits are acted upon by the pink surfaces representing
Z-stabilizers and green lines/surfaces indicating X-stabilizers. \cite{chatterjee2025q}.}
    \label{fig:surfacecode}
\end{figure}

Surface codes are a class of topological quantum error correction codes (QECCs) that encode logical qubits into a two-dimensional lattice of physical qubits \cite{fowler2012,dennis2002}. They belong to the broader family of stabilizer codes, which use a set of commuting Pauli operators called stabilizer generators to define the code space. These stabilizers are measured periodically to detect and correct errors without collapsing the quantum state. Each stabilizer is either an \( X \)-type or \( Z \)-type operator acting on  small group of qubits. \( X \)-type stabilizers detect phase-flip (\( Z \)) errors, and \( Z \)-type stabilizers detect bit-flip (\( X \)) errors. The syndrome, i.e., the set of measurement outcomes from these stabilizers, helps identify error locations. Ancilla (syndrome) qubits are used to perform these measurements via circuits involving CNOT or CZ gates, preserving the coherence of the data qubits. Unrotated \cite{fowler2012} and rotated \cite{tomita2014low} surface codes are two different layouts for implementing surface code QEC (Fig. \ref{fig:surfacecode}).

In an unrotated surface code layout, data qubits lie on the edges of a square lattice, and each face (plaquette) is associated with a stabilizer. Vertices host \( X \)-type stabilizers and faces host \( Z \)-type stabilizers. In contrast, rotated surface codes optimize the layout by tilting the lattice 45 degrees and using fewer physical qubits and ancillae for the same code distance \cite{siegel2023adaptive, chatterjee2025q}. Rotated codes simplify implementation and improve error thresholds. Let \( d \) denote the code distance. An unrotated code of distance \( d \) typically requires \( (2d^2 - 1) \) qubits, while a rotated layout uses around \( d^2 \) qubits, providing a more compact structure. Hence, in this work, we focus exclusively on rotated surface codes.

\subsection{Physical vs. Logical Errors}

In quantum computing, on one hand, physical errors refer to imperfections in the operations of individual qubits, such as decoherence, gate infidelity, and measurement errors. These errors are inherent to the hardware and can vary across qubits and over time. Logical errors, on the other hand, are errors that occur at the level of the encoded logical qubit. The goal of QEC is to suppress logical errors by detecting and correcting physical errors before they accumulate.

The performance of a surface code is characterized by its code distance \( d \), which determines the number of physical errors the code can correct. A distance-\( d \) code can correct up to \( \lfloor (d-1)/2 \rfloor \) errors. The logical error rate \( p_L \) typically decreases exponentially with increasing \( d \), assuming the physical error rate \( p \) is below a certain threshold \( p_{\mathrm{th}} \) \cite{fowler2012, terhal2015}:
\begin{equation}
    p_L \approx \alpha \left( \frac{p}{p_{\mathrm{th}}} \right)^{(d+1)/2},
\end{equation}
where \( \alpha \) is a constant dependent on the decoder and code layout.



\subsection{Adaptive Quantum Error Correction and Related Works}

Adaptive QEC strategies modify the error correction process to match the specific noise characteristics of a quantum system at a given time. Several approaches have been proposed, including noise-adapted recovery circuits such as the Petz map and its generalizations, which aim to optimize the recovery map for a known noise channel \cite{jayashankar2023quantum}. These methods often use average or worst-case fidelity as optimization objectives and are sometimes formulated as semidefinite programs for practical implementation. Most existing adaptive methods focus on tuning the decoding or recovery operations without changing the encoding structure, such as the code distance. For example, recovery maps tailored to amplitude damping noise offer near-optimal performance in that regime, but assume a fixed codespace and do not address variations in hardware noise across time and space. Additionally, many of these methods are designed for specific error models and may not generalize to more complex or time-varying noise profiles observed in real systems.

In contrast, we introduce a dynamic code distance selection strategy that adjusts surface code parameters based on per-qubit error rates obtained from daily hardware calibration data. This enables both encoding and decoding to respond to real hardware conditions. Unlike prior techniques that assume static encoding and focus on decoder optimization, we treat QEC configuration as a resource allocation problem that balances logical error suppression with physical qubit overhead.

\section{Error Characterization and Code Distance Estimation}

\subsection{Pauli-X and CNOT Error Variation in Real Devices}

We use calibration data from the \texttt{ibm\_kyiv} superconducting quantum processor, which consists of 127 transmon qubits arranged in a heavy-hex lattice. The dataset includes daily calibration results collected over 12 different days. Each calibration snapshot reports various hardware metrics such as coherence times, readout assignment errors, and gate-level error rates. For this study, we focus on single-qubit Pauli-X gate error rates and two-qubit CNOT gate error rates.

\begin{figure}[ht]
    \centering
    \includegraphics[width=\linewidth]{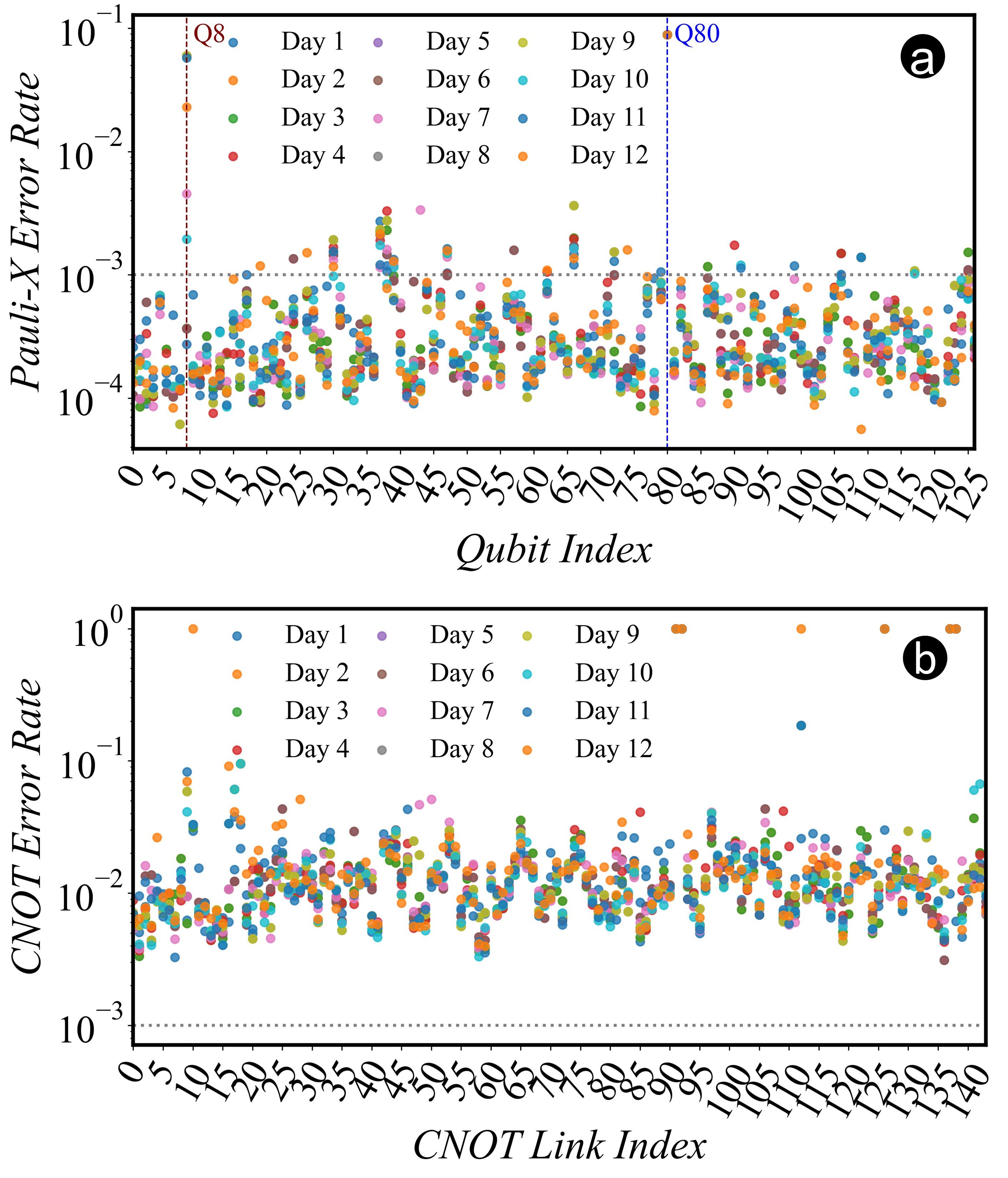}
    \caption{(a) Pauli-X error rate across 12 calibration days for all 127 qubits in the \texttt{ibm\_kyiv} processor. Each dot represents the Pauli-X error of one qubit per day. Qubit 8 (red dashed line) shows large temporal variation, while Qubit 80 (blue dashed line) maintains consistently high error. A dotted horizontal line marks the reference threshold at $10^{-3}$. (b) CNOT error rate across 12 calibration days for all 144 CNOT links. Each dot represents the CNOT error of one link per day. Link indices are assigned arbitrarily based on their order in the calibration data.}
    \label{fig:pauli_and_cnot}
\end{figure}

The Pauli-X error rates are extracted for all qubits from each of the 12 calibration files. These values reflect the combined effect of gate fidelity, crosstalk, and temporal fluctuations in control electronics. To visualize this variation, we plot the Pauli-X error rate for each qubit across the 12 days (Fig.~\ref{fig:pauli_and_cnot}(a)). The figure reveals key patterns. On one hand, some qubits such as Qubit 8, show large fluctuations in error rate across days. Others, like Qubit 80, consistently exhibit high error rates well above the practical threshold for error correction. On the other hand, many qubits remain within a low and stable error range, indicating they are better candidates for low-overhead encoding. This observation motivates the need for a dynamic code distance strategy that responds to current hardware conditions rather than using a fixed configuration for all qubits.

We also observe significant variability in two-qubit gate performance. Fig.~\ref{fig:pauli_and_cnot}(b) shows the CNOT error rate over 12 days for all 144 unique CNOT links in the \texttt{ibm\_kyiv} device. Each link is indexed and plotted independently. The majority of CNOT links show error rates ranging from $10^{-2}$ to $10^{-1}$, with several links occasionally exceeding $10^{-1}$. A few links exhibit abrupt spikes on certain days, indicating strong temporal fluctuations. Since CNOT operations are used heavily in stabilizer measurement and syndrome extraction circuits, such variation can directly impact the effectiveness of surface code decoding. In this paper we restrict our analysis and logical error simulations to only single-qubit Pauli-X error rates. This is because Pauli-X errors offer a straightforward way to study how physical errors translate to logical errors. Including CNOT noise would require a larger simulation space and assumptions about correlated error behavior, which we leave for future work.

\subsection{Logical Error Estimation and Distance Selection}

To evaluate how physical error variation impacts logical reliability, we simulate the logical error rate as a function of physical error rate across a range of surface code distances. This analysis helps identify the code distance necessary to maintain a target logical error rate.

\begin{figure}[ht]
    \centering
    \includegraphics[width=0.9\linewidth]{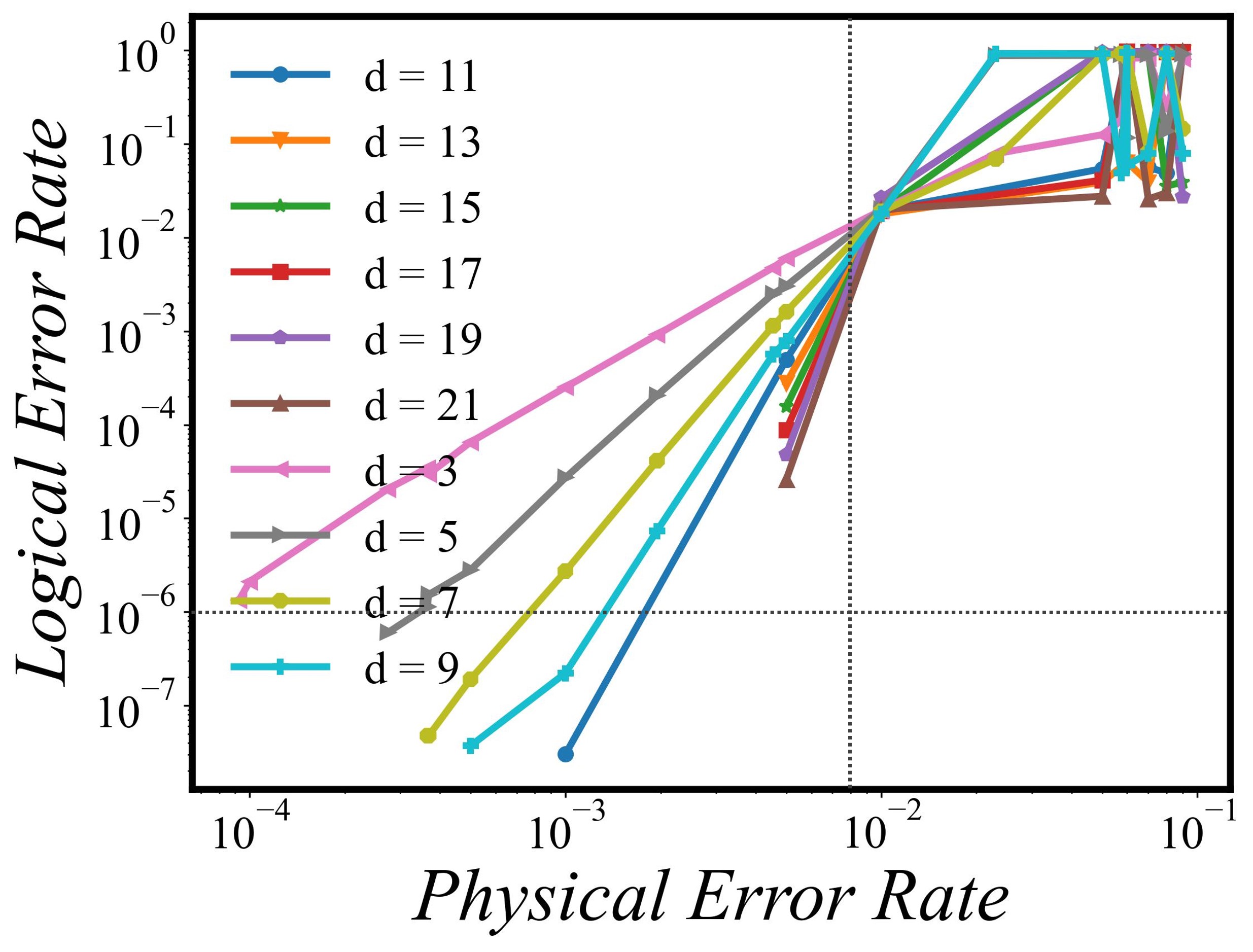}
    \caption{Logical error rate as a function of physical Pauli-X error rate for surface code distances ranging from 3 to 21. The horizontal dashed line shows a target logical error rate of $10^{-6}$. The vertical dashed line marks the cutoff physical error rate above which QEC is ineffective.}
    \label{fig:logical_vs_physical}
\end{figure}

Fig.~\ref{fig:logical_vs_physical} shows the logical error rate vs. physical error rate for distances ranging from \( d = 3 \) to \( d = 21 \). Each line corresponds to a different code distance. The x-axis covers the full range of Pauli-X error rates observed across the 127 qubits in the \texttt{ibm\_kyiv} device over 12 days. The y-axis shows the corresponding logical error rate for a rotated surface code simulated using the open-source simulator \texttt{Stim}. This tool is well-suited for large-scale quantum error correction simulations and accurately captures fault-tolerant behavior under realistic error models, as demonstrated in prior work \cite{chatterjee2025q}. In our experiments, we use a circuit-level noise model with symmetric depolarizing errors applied before and after Clifford gates, resets, and measurements. Each logical circuit uses a code distance \( d \)  and is repeated for \( 3d \)  rounds. Decoding is performed using the minimum-weight perfect matching decoder \texttt{PyMatching}.

The figure shows that logical error rate drops exponentially with increasing code distance as expected, but this improvement only holds below a certain physical error threshold. We include a horizontal dashed line at a logical error rate of $10^{-6}$. This value reflects a common design target for practical fault-tolerant quantum computing~\cite{fowler2012}. Several large-scale system proposals adopt this threshold to ensure reliable execution of deep quantum circuits with surface code protection. A vertical dashed line indicates the physical error rate at which this logical target becomes unreachable with reasonable distances. In our simulation, this cutoff occurs near $p \approx 8 \times 10^{-3}$. Qubits with higher error rates would require prohibitively large distances, which is not realistic on current NISQ hardware.

This result supports our approach: rather than using a fixed distance for all qubits, we propose selecting distances based on each qubit's current error rate. Stable qubits can be encoded with smaller distances, saving resources. Qubits that exceed the cutoff can be excluded from encoding or replaced with lower-error alternatives, leading to a more resource-efficient and reliable QEC strategy.

\section{QEC Optimization Strategy}


Our goal is to optimize resources without sacrificing reliability. A large distance provides stronger error suppression, but it also increases physical qubit overhead quadratically. For example, a distance-9 rotated surface code requires 81 physical qubits per logical qubit. A distance-13 code already needs 169. Using the same large code for every qubit would waste resources, especially for qubits that are already stable.

\subsection{Methodology}

We propose following procedure for adaptive QEC configuration:

\begin{itemize}
    \item \textbf{Collect error rates} from daily calibration data.
    \item \textbf{Sort all qubits} on a given day by their error rates.
    \item \textbf{Determine required distance} for each qubit to meet the logical error threshold (e.g., \(10^{-6}\)), using simulation data.
    \item \textbf{Exclude high-error qubits} that need distances beyond the allowed maximum (e.g., distance \(> 9\)).
    \item \textbf{Assign each qubit the smallest distance} that keeps the logical error below the target, only if such a distance is possible.
\end{itemize}

This process is repeated for each calibration snapshot. It enables distance assignments to track qubit performance over time. This strategy has three advantages. First, it avoids unnecessary overhead for low-error qubits. Second, it prevents unreliable encoding of high-error qubits. Third, it allows a flexible, day-by-day adjustment of the system’s logical qubit capacity, depending on the number of usable physical qubits.

\subsection{Results and Trade-offs}

\begin{figure}[t]
    \centering
    \includegraphics[width=0.95\linewidth]{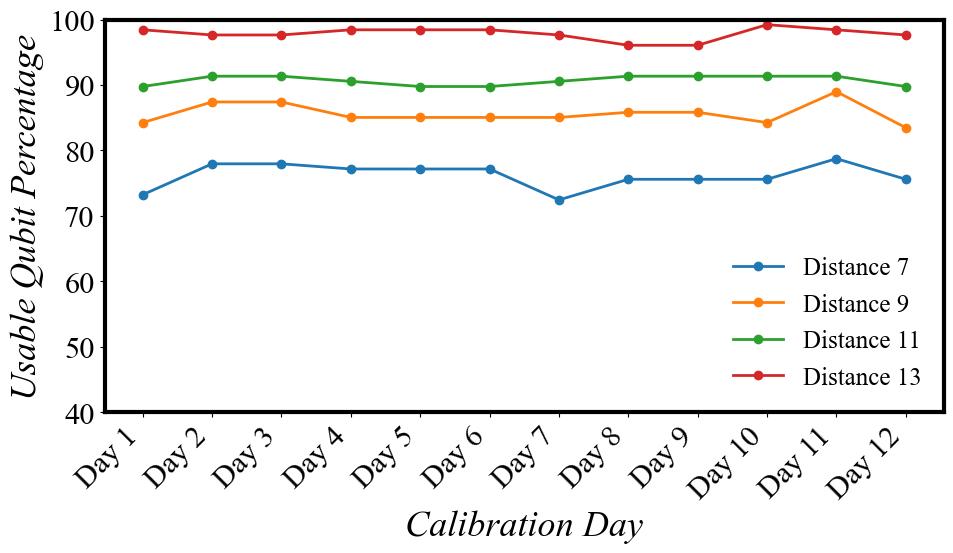}
    \caption{Percentage of usable qubits on each calibration day across four code distances. A qubit is considered usable if its Pauli-X \emph{physical} error rate is below the threshold required to achieve a target \emph{logical} error rate of $10^{-6}$ at that code distance. These threshold values, obtained from simulation, are: $7\times10^{-4}$ for distance-7, $10^{-3}$ for distance-9, $2\times10^{-3}$ for distance-11, and $7\times10^{-3}$ for distance-13. Higher distances tolerate more physical error, allowing more qubits to be used, but increase resource cost.
    }
    \label{fig:usable_qubit_lineplot}
\end{figure}


We apply the proposed strategy to 12 days of calibration data from the \texttt{ibm\_kyiv} device. Fig.~\ref{fig:usable_qubit_lineplot} shows the percentage of qubits usable on each of the 12 calibration days under four different code distances: 7, 9, 11, and 13. A qubit is considered usable if its Pauli-X error rate is below the threshold required to achieve a target logical error rate of $10^{-6}$. These thresholds, obtained from Fig. ~\ref{fig:logical_vs_physical}, are $7\times10^{-4}$ for distance-7, $10^{-3}$ for distance-9, $2\times10^{-3}$ for distance-11, and $7\times10^{-3}$ for distance-13. 

The figure reveals that higher distances tolerate higher physical error, resulting in a larger fraction of usable qubits. On average, only about 76\% of qubits are usable with distance-7. This increases to approximately 86\% with distance-9, around 91\% for distance-11, and nearly 98\% for distance-13. These results show a trade-off between error tolerance and resource cost. Higher code distances allow more qubits to be used but require more physical qubits per logical one. This result highlights the importance of adaptive QEC, where distance is chosen based on each qubit's quality. It improves hardware usage without unnecessary overhead.

To understand the resource benefit of adaptive QEC, we compare two scenarios:

\begin{itemize}
    \item \textbf{Baseline QEC:} A single worst-case error rate across all qubits and days would force us to design for the maximum code distance needed, in this case, distance 13. This requires $13^2 = 169$ physical qubits per logical qubit.
    
    \item \textbf{Adaptive QEC:} We exclude high-error qubits and use smaller distances when possible. For instance, with a maximum distance of 9, we can still use approximately 85\% of the qubits (about 108 out of 127). Each logical qubit then uses only $9^2 = 81$ physical qubits.
\end{itemize}

This leads to a resource saving of roughly 52\% per logical qubit. Moreover, the number of usable qubits remains high enough to support practical encoding even with the stricter constraint. Therefore, we achieve a balance between resource efficiency and fault tolerance by adjusting the code distance based on qubit quality.

\begin{figure}[t]
    \centering
    \includegraphics[width=0.95\linewidth]{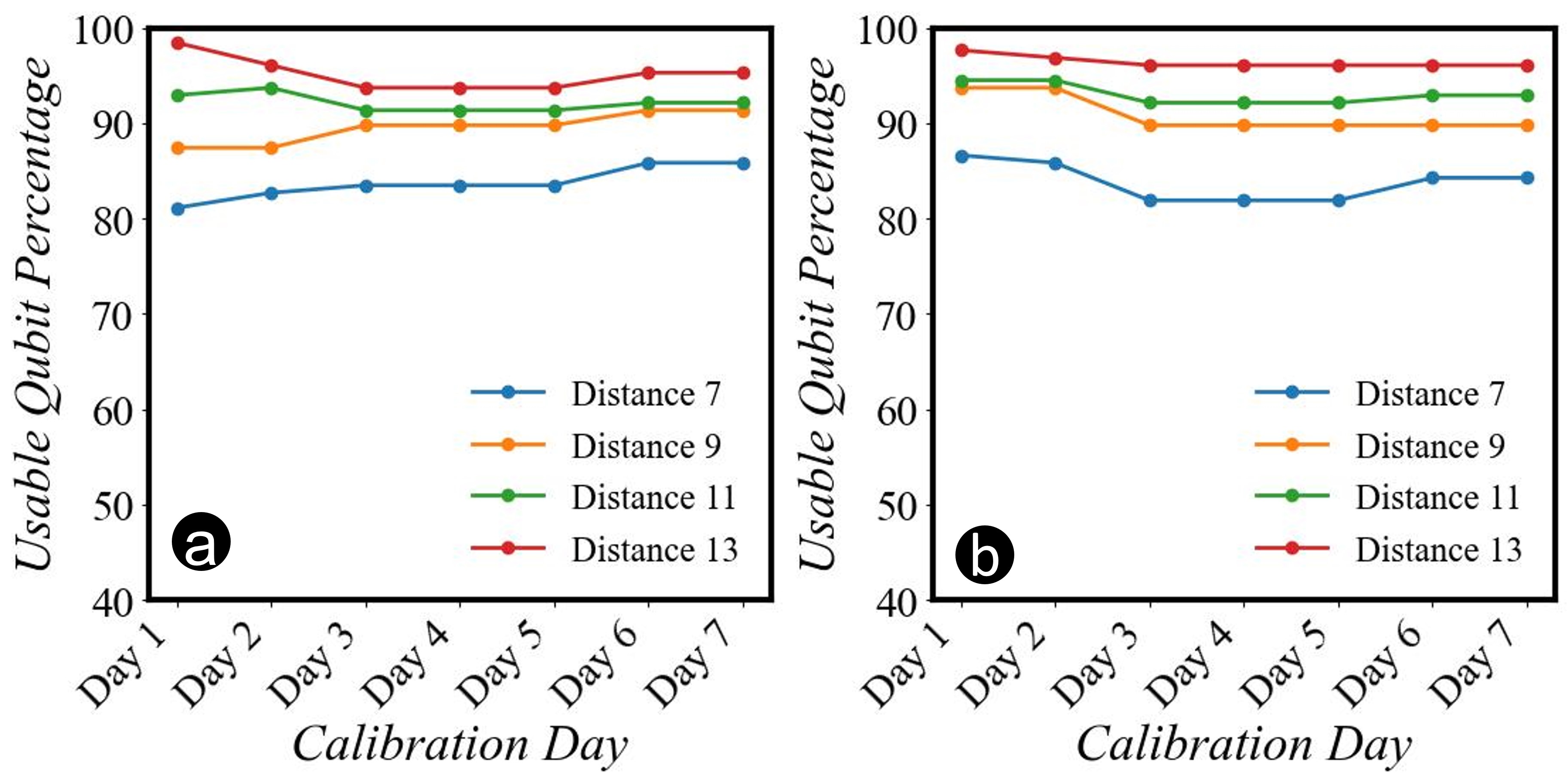}
    \caption{Percentage of usable qubits across seven calibration days for two additional 127-qubit devices: (a) \texttt{ibm\_brisbane} and (b) \texttt{ibm\_sherbrooke}. For both devices, average usability improves from about 82\% at distance-7 to over 98\% at distance-13.}
    \label{fig:brisbane_sherbrooke}
\end{figure}

We validate our method on two more 127-qubit devices, \texttt{ibm\_brisbane} and \texttt{ibm\_sherbrooke}, using 7 days of calibration data. Fig.~\ref{fig:brisbane_sherbrooke} shows the usable qubit percentage on each day for four fixed code distances. Similar to \texttt{ibm\_kyiv}, the results show a consistent trend: higher code distances allow more qubits to be used. For example, on \texttt{ibm\_brisbane}, using distance-7 yields around 81--85\% usable qubits, while distance-13 consistently enables over 97\%. On \texttt{ibm\_sherbrooke}, distance-7 usage stays above 82\%, and distance-13 again exceeds 98\% on all days. Because over 80\% of the qubits remain usable even with distance-7, we can consider using this stricter distance for these devices. Compared to distance-13, which requires 169 physical qubits per logical qubit, a distance-7 code needs only 49. This results in a $\sim$71\% reduction in overhead, while still maintaining high hardware utilization. These results confirm that adaptive QEC can significantly reduce resource cost. By excluding a small number of high-error qubits and assigning code distances according to error level, we achieve more efficient QEC without compromising fault tolerance.

\subsection{Implications and Limitations}

This method also supports hardware-aware compilation. By knowing in advance which qubits are available for encoding and what distance to use, a compiler or scheduler can better map logical circuits to hardware. However, there are also trade-offs. This approach assumes calibration data is reliable and available before execution. It also assumes we can adjust decoder behavior to match the changing distances.

Despite these challenges, our results show that adaptive code assignment provides a more realistic and efficient way to deploy QEC on NISQ hardware. We avoid the inefficiency of one-size-fits-all encoding and instead match protection level to actual qubit quality. This framework can also be extended in the future to support noise bias, correlated errors, and decoding-aware layout optimization.

\section{Conclusion}
\label{sec:conclusion}

This paper presents a practical method for adapting quantum error correction codes based on the variation in physical qubit quality. Using calibration data from IBM's 127-qubit superconducting device, we show that both single-qubit and two-qubit error rates vary significantly across qubits and over time. We evaluate the impact of these variations on logical error rates and demonstrate that a fixed code distance is not efficient or reliable across all qubits. To address this, we propose an adaptive approach that selects code distance based on daily error rates. High-error qubits are excluded if they require distances beyond a practical limit. This method reduces overhead and increases the number of usable qubits without sacrificing logical fidelity. Our analysis shows that adjusting the code based on real error data allows more efficient use of quantum hardware, especially in NISQ systems where resources are limited.

\section*{Acknowledgements}
This work is supported in parts by NSF (CNS-1722557, CNS-2129675, CCF-2210963, CCF-1718474, OIA-2040667, DGE-1723687, DGE-1821766, and DGE-2113839), Intel’s gift and seed grants from Penn State ICDS and Huck Institute of the Life Sciences. 

\bibliography{Manuscript}

\end{document}